\documentclass[preprint,12pt]{elsarticle}

\usepackage{graphicx}
\usepackage{subfig}
\usepackage{bm}
\usepackage{amssymb}
\usepackage{url}
\usepackage[strings]{underscore}

\usepackage{xcolor}
\usepackage[printwatermark]{xwatermark}

\journal{Computer Physics Communications}

\begin{document}

\begin{frontmatter}

\title{MIST: A Simple and Efficient Molecular Dynamics Abstraction Library for Integrator Development}

\author[a,f]{Iain Bethune\corref{author}}
\author[b]{Ralf Banisch}
\author[c]{Elena Breitmoser}
\author[c]{Antonia B. K. Collis}
\author[c]{Gordon Gibb}
\author[d]{Gianpaolo Gobbo}
\author[e]{Charles Matthews}
\author[f]{Graeme J. Ackland}
\author[g]{Benedict J. Leimkuhler}

\cortext[author] {Corresponding author.\\\textit{E-mail address:} iain.bethune@stfc.ac.uk}
\address[a]{STFC Hartree Centre, Sci-Tech Daresbury, Warrington, WA7 6UE, UK}
\address[b]{Department of Mathematics and Computer Science, Freie Universit{\"a}t Berlin, Arnimallee 6, 14195 Berlin, Germany}
\address[c]{EPCC, The University of Edinburgh, James Clerk Maxwell Building, Peter Guthrie Tait Road, Edinburgh, EH9 3FD, UK}
\address[d]{Department of Chemical Engineering, Massachusetts Institute of Technology, 77 Massachusetts Avenue, Cambridge, MA 02139, USA}
\address[e]{Department of Statistics, University of Chicago, S. Ellis Avenue, Chicago, IL 60637, USA}
\address[f]{School of Physics and Astronomy, The University of Edinburgh, James Clerk Maxwell Building, Peter Guthrie Tait Road, Edinburgh, EH9 3FD}
\address[g]{School of Mathematics, The University of Edinburgh, James Clerk Maxwell Building, Peter Guthrie Tait Road, Edinburgh, EH9 3FD}

\begin{abstract}
We present MIST, the Molecular Integration Simulation Toolkit, a lightweight and efficient software library written in C++ which provides an abstract interface to common molecular dynamics codes, enabling rapid and portable development of new integration schemes for molecular dynamics.  The initial release provides plug-in interfaces to NAMD-Lite, GROMACS and Amber, and includes several standard integration schemes, a constraint solver, temperature control using Langevin Dynamics, and two tempering schemes.  We describe the architecture and functionality of the library and the C and Fortran APIs which can be used to interface additional MD codes to MIST.

We show, for a range of test systems, that MIST introduces negligible overheads for serial, shared-memory parallel, and GPU-accelerated cases, except for Amber where the native integrators run directly on the GPU itself.  As a demonstration of the capabilities of MIST, we describe a simulated tempering simulation used to study the free energy landscape of Alanine-12 in both vacuum and detailed solvent conditions.


\end{abstract}


\end{frontmatter}


{\bf PROGRAM SUMMARY}

\begin{small}
\noindent
{\em Program Title:} MIST - Molecular Integration Simulation Toolkit       \\
{\em Licensing provisions:} BSD 2-clause                                   \\
{\em Programming language:} C++ (C and Fortran interfaces)                 \\
{\em Nature of problem:}\\
 Production Molecular Dynamics codes have become increasingly complex, making it difficult to implement new functionality, especially algorithms that modify the core MD integration loop.  This places a barrier in the way of new algorithms making their way from theory to implementation. \\
{\em Solution method:}\\
 MIST provides a simplified abstract interface for integrator developers that may be interfaced via source-code patches and a library API to a variety of MD codes, with minimal loss of performance. \\
{\em Restrictions and Unusual features:}\\
 MIST interfaces only to specific versions of MD codes: Amber 14, Gromacs 5.0.2 and NAMD-Lite 2.0.3\\
{\em Comments:}\\
 MIST is freely available from \url{https://bitbucket.org/extasy-project/mist}.
\end{small}

\section{Background}

Molecular Dynamics with classical force-fields has proved to be an extraordinarily successful method for studying dynamical processes as well as sampling the conformational space of complex macromolecules (see \cite{elber2016} for a recent review).  This success is largely due to advances in four directions; improving accuracy of force-fields, developing faster and more scalable force calculations,  increasing computational power of high performance computing systems, and advanced sampling algorithms such as metadynamics~\cite{Laio2002}, replica-exchange MD~\cite{Sugita1999} and parallel tempering~\cite{Hansmann1997}.  A number of highly-optimised MD packages such as NAMD~\cite{NAMD} and GROMACS~\cite{GROMACS} have been developed which implement a range of different force-fields, are able to run on a range of commodity (x86 CPU clusters and GPUs) and special-purpose~\cite{Shaw2008} hardware and represent many hundreds of person-years of effort.  All of this functionality and performance comes at a cost in terms of code complexity, and even if an MD code is open-source, in practice it is difficult for developers to add significant new features without close collaboration with the main developers of the code.

The result is that the core algorithms used for MD timestepping evolve slowly. Typical integration schemes are based on velocity Verlet or leapfrog integration, combined with one of several common thermo- or baro-stats~\cite{Berendsen1984,Hoover1985,Martyna1992,Parrinello1980}.   Recent innovation has centered on higher level methods for promoting space exploration~\cite{Zheng2013} or modifying the potential energy surface to lower barriers between metastable states~\cite{Hamelberg2004}.  We argue that there is ``room at the bottom''\footnote{With apologies to Richard P. Feynman.} for innovative methods which modify the core integration step to access larger time steps and/or improved sampling accuracy (e.g. \cite{Leimkuhler2013, Gobbo2015, Dullweber1997, Leimkuhler2016, Ceriotti2010, Morrone2011}) which have not yet been implemented in any `production' MD codes.

The \emph{status quo} is a catch-22 for applied mathematicians: if new algorithms cannot be easily incorporated into widely used MD packages, then it is impossible to demonstrate their benefits on complex systems of practical interest.  If such demonstrations are not available, there will be little interest from the MD user community, and there is no incentive for MD package developers to implement the methods; in many cases algorithms are left `on the shelf' for long periods.

Our solution to this conundrum is MIST--the Molecular Integration Simulation Toolkit.  MIST is a software library (available from \url{https://bitbucket.org/extasy-project/mist}) which can be easily interfaced to a variety of MD codes (currently GROMACS, Amber~\cite{Case2005}, NAMD-Lite~\cite{namdlite2007} and Forcite~\cite{Forcite}) and which provides an abstract interface to the state of the system.  This enables integration algorithms to be programmed without concern for the complexities of a typical MD code, with low performance overhead.  We describe the architecture of the MIST library, the Application Programmer's Interface (API) which plugs in to existing codes, explain how to implement an integrator, and illustrate the performance on a range of computational benchmark tests.

Although we use the term ``integrator'' here to describe timestepping procedures used in MD, MIST is deliberately not  restrictive regarding the types of equations that can be simulated; indeed the example we present at the end of this article implements simulated tempering, a complex enhanced sampling strategy, within MIST.  While MIST currently supports a range of classical MD codes, the design of MIST is flexible enough that it could also be used for \emph{ab initio} MD based on the Born-Oppenheimer approximation.

\subsection{Related Work}

PLUMED\cite{Tribello2014}, is similar in design to MIST in that it is a software library that interfaces to a range of MD codes via API calls (which may be inserted using source-code patches).  While PLUMED is widely used and at present supports more MD codes than MIST, it only allows the calculated MD forces to be modified and does not provide read/write access to the atomic positions and velocities.  Thus, it is less flexible than MIST and facilitates a much smaller set of MD algorithms.

A simplified MD program (such as NAMD-Lite removes much of the complexity of a production MD code, making it easy to modify.  However, this results in a loss of functionality (e.g. forcefield support, analysis tools, properties calculations) and performance--restricting the scale of problems which can be tackled.

OpenMM~\cite{Eastman2017} is a toolbox for building MD applications which is designed to be extensible at the source-code level, while being portable to a range of CPU and GPU hardware.  The \texttt{CustomIntegrator} interface is flexible and provides a Python API to allow declaration of (for example) variables which should be computed for each degree of freedom.  However, we argue that this API approach results in code which is less clear and intuitive than the way an integrator is specified in MIST.  Moreover, the idea in MIST is that the mathematical integration algorithm framework is independent of the MD engine (which encompasses force-field evaluation, boundary conditions, etc.); this allows comparison and cross-validation of results using alternative molecular software systems.  Ultimately MIST offers improved interoperability with a broad range of existing codes and force-fields.

\section{MIST Architecture}

\begin{figure}
\centering{
\includegraphics[scale=0.6]{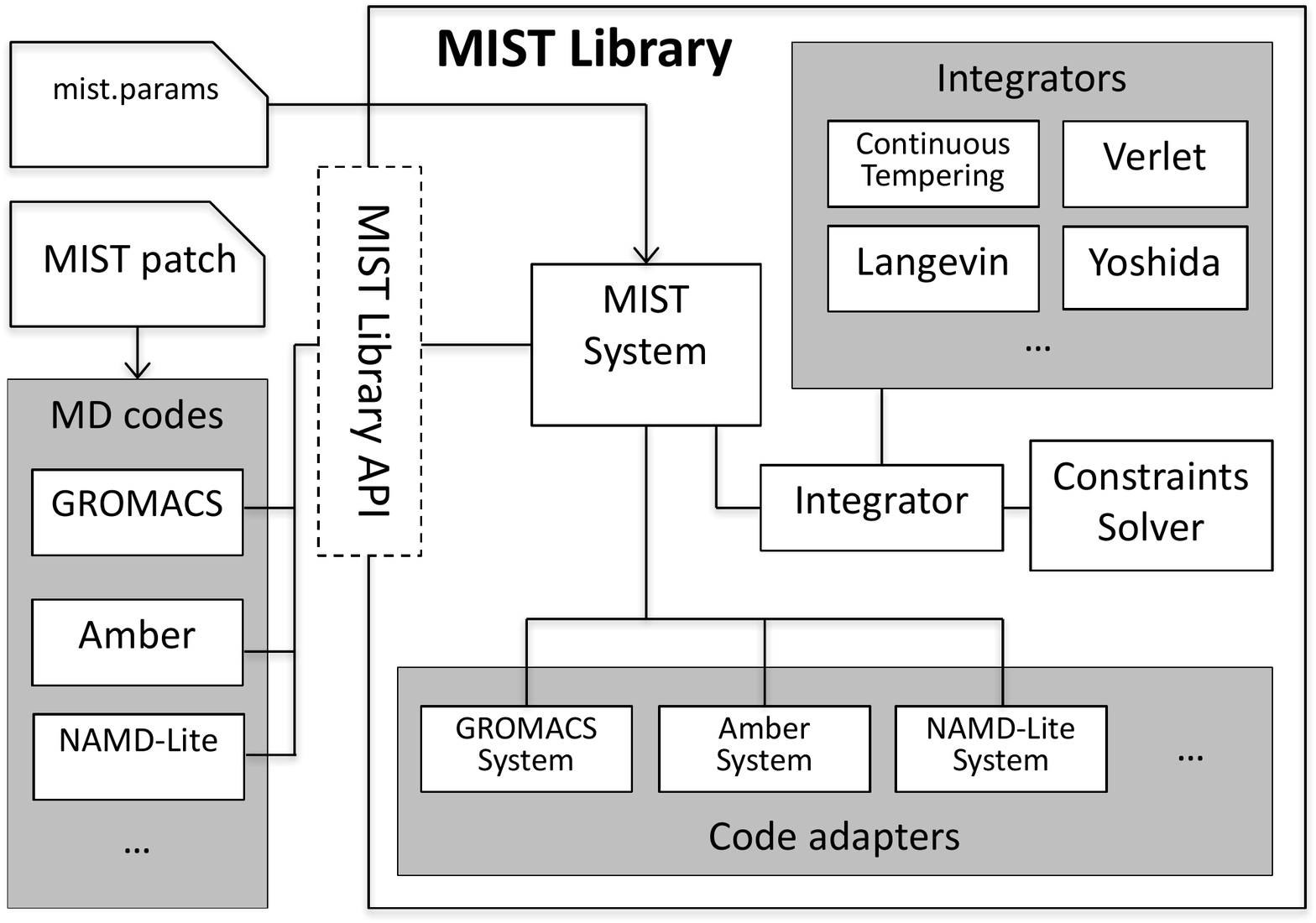}
\caption{Schematic representation of the main components of the MIST library.}
\label{fig:mist-arch}
}
\end{figure}

The design of MIST balances three main objectives: providing an abstraction of the state of a molecular dynamics simulation which is sufficiently general to allow it to be implemented in conjunction with any particular MD code; implementing this interface with a low performance overhead relative to standard MD codes; and providing an intuitive and expressive interface for developing integrators.  Figure~\ref{fig:mist-arch} shows how the components of the library are related.  A `host' MD code makes use of the MIST library to perform time integration of the state of the system.  At the core of the MIST library is a representation of the state of the system (and a set of adaptors for specific MD codes), which can be used by developers to implement new integrator algorithms.  Each of these components is discussed in detail in the following sections.  Several integrators are included in the library, to serve as templates, as well as to provide new capabilities to users (see Table~\ref{table:integrators}).

\subsection{The MIST System}
\label{sec:mist-system}

The conceptual model of the state of the system is very simple.  We have a set of $n$ point particles labeled $0..n-1$, where $n$ is assumed to remain fixed for the duration of the simulation.  Each particle has a set of properties: position, velocity, mass, kind (atomic species, typically) and force (the force acting on the particle).  In addition, there are a number of global properties, such as the cell lattice vectors (if periodic boundary conditions are employed), the total potential energy.  This state is encapsulated as a C++ \texttt{System} class.  For the dynamical variables (position and velocity), accessor (e.g. \texttt{GetPosition()}) and mutator (e.g. \texttt{SetPosition()}  methods are provided.  The other properties are read-only, and so only accessors are provided (e.g. \texttt{GetForce()}). Evaluation of forces is treated as a black-box, and a method \texttt{UpdateForces()} is provided to request the forces on each particle to be updated (usually the most expensive operation in an MD simulation).  Access to a simple representation of the molecular topology is also provided: a set of $b$ bonds labelled $0..b-1$, where each bond consists of a pair of particle indices and a fixed length, encapsulated as a lightweight \texttt{Bond} object.  An important aspect of this model is that in this release of MIST we require that all of the data is stored within a single address space i.e. domain decomposition using MPI is not supported (although this is under development, see Section~\ref{sec:developments}).  Shared memory parallelisation such as OpenMP is permitted in the host code, and indeed is used within MIST.

The \texttt{System} provides everything that is needed to implement an integrator (see Section~\ref{sec:integrators}) but requires an adaptor to implement the MD-code-independent \texttt{System} methods using the data structures present in a particular MD code.  The choice of MD code is made at compile-time via arguments to the \texttt{configure} script used to drive MIST's build process.  For simplicity and performance, we provide MIST with access to the raw data structures in an MD code through pointers registered with MIST by the host MD code.  This allows the library API (see Section~\ref{sec:mist-api}) to be remain completely code-agnostic, and the details of how those pointers are interpreted to yield useful data is encapsulated with the code-specific \texttt{System} adaptor classes.  For example, GROMACS (by default) stores data as arrays of single-precision floating point whereas Amber and NAMD-Lite use double-precision, and GROMACS stores the inverse masses of particles, rather than the masses themselves. These differences are hidden from the user by the \texttt{System} abstraction.

\subsection{Integrators}
\label{sec:integrators}

In MIST, an \texttt{Integrator} is an abstract class which has a single method which must be implemented by any sub-class: \texttt{void Step(double dt)}, as the name suggests, a function which implements the time integration of the system state from $t$ to $t+dt$, according to some algorithm.  To add a new integration algorithm to the library, a developer need only  create a new class which inherits from \texttt{Integrator} and implements the \texttt{Step} method.  A number of convenience functions are also included in the base class which simplify coding, for example a velocity Verlet integrator for the NVE ensemble is as simple as:

\begin{verbatim}
void VerletIntegrator::Step(double dt)
{
    // Velocity half-step
    VelocityStep(0.5 * dt);

    // Position full step
    PositionStep(dt);

    system->UpdateForces();

    // Velocity half-step
    VelocityStep(0.5 * dt);
}
\end{verbatim}

More complex algorithms can be implemented by directly updating individual particle properties using methods of the \texttt{System} class.  For example, the stochastic part of our Langevin dynamics integrator is implemented as (\texttt{c1} and \texttt{c3} are double precision floating-point constants, \texttt{v} is a variable of the lightweight \texttt{Vector3} type, and \texttt{rnd[tid]} is a (thread-local) instance of our random number generator):

\begin{verbatim}
for (int i = 0; i < system->GetNumParticles(); i++)
{
    v = system->GetVelocity(i);
    sqrtinvm = system->GetInverseSqrtMass(i); // 1/sqrt(m)
    v.x = c1 * v.x + sqrtinvm * c3 * rnd[tid]->random_gaussian();
    v.y = c1 * v.y + sqrtinvm * c3 * rnd[tid]->random_gaussian();
    v.z = c1 * v.z + sqrtinvm * c3 * rnd[tid]->random_gaussian();
    system->SetVelocity(i, v);
}
\end{verbatim}

Most integrators advance the entire state of the system from time $t$ to $t+dt$.  However, a class of algorithms such as Verlet integration in the `leapfrog' formulation operate assuming the velocities to be offset by $dt/2$ from the positions at the start of each step.  In MIST, such an integrator must be labeled with the `feature flag' \texttt{MIST\_FEATURE\_POS\_VEL\_OFFSET\_PLUS\_HALF\_DT}, to ensure that account of this is taken by the host MD code (for example, computing kinetic energy based on averages over two steps).  The feature flags are used to signal any special requirements that the integrator might place on the host code, for example that access to individual components of the force-field is required (\texttt{MIST\_FEATURE\_FORCE\_COMPONENTS}).  This allows integrators to be coded quite generally and functionality be added incrementally to the MD code adaptors, with a check performed at startup to see if the features of the selected integrator are supported by the code.

Selecting an integrator and setting parameters are done through an input file \texttt{mist.params} which contains a list of (case-insensitive) key-value pairs.  A separate file is used so that integrator settings are code-independent, whereas run control settings such as the number of time steps, when to write trajectory output etc. are managed by the usual input file(s) of the host code.  A minimal input file to use velocity Verlet integration in the NVE ensemble would be:

\begin{verbatim}
integrator verlet
\end{verbatim}

A slightly more complex example, to run Langevin NVT dynamics at 300K would be:

\begin{verbatim}
integrator langevin
langtemp 300 # Target temperature, in K
langfriction 1.0 # Friction parameter gamma, in ps^(-1)
\end{verbatim}

A full list of integrators and possible parameters (including default values and units) is available online at \url{https://bitbucket.org/extasy-project/mist/wiki/MIST%20Integrators} and is summarised in Table \ref{table:integrators}.  Since the lattice parameters are fixed in this release of MIST rather than treated as dynamical variables, at present it is not possible to implement constant-pressure MD schemes - this is under development for a future release (see Section~{\ref{sec:developments}}).

\begin{table}
\footnotesize
\centering{
\begin{tabular}{| l | p{1.8cm} | l | p{6cm} |}
\hline
Keyword & Supports Constraints & Ensemble & Description \\
\hline
\texttt{verlet} & Yes & NVE & Velocity Verlet \\
\texttt{leapfrog} & Yes & NVE & Verlet Leapfrog \\
\texttt{langevin} & Yes & NVT & Langevin Dynamics with BAOAB or ABOBA splitting method \cite{Leimkuhler2013} \\
\texttt{yoshida4} & No & NVE & Symplectic 4th order integrator \cite{Yoshida1990} \\
\texttt{yoshida8} & No & NVE & Symplectic 8th order integrator \cite{Yoshida1990} \\
\texttt{tempering} & Yes & - & Continuous Tempering - enhanced sampling with continously varying temperature \cite{Gobbo2015} \\
\texttt{tamd} & No & - & Temperature Accelerated MD - Langevin dynamics with an additional thermostat coupled to two backbone dihedral angles \\
\texttt{tempering_tamd} & No & - & TAMD where additional degrees of freedom are heated via the Continuous Tempering algorithm \\
\texttt{simulated_tempering} & Yes & - & Simulated Tempering with on-the-fly weight determination~\cite{Nguyen2013,Zhang2015} using Langevin thermostat \\
\hline
\end{tabular}
\caption{List of current MIST integrators and algorithms}
 \label{table:integrators}
}
\end{table}

\subsection{MIST Library API}
\label{sec:mist-api}

\begin{figure}
\vspace{-3cm}
\centering{
\includegraphics[scale=1.8]{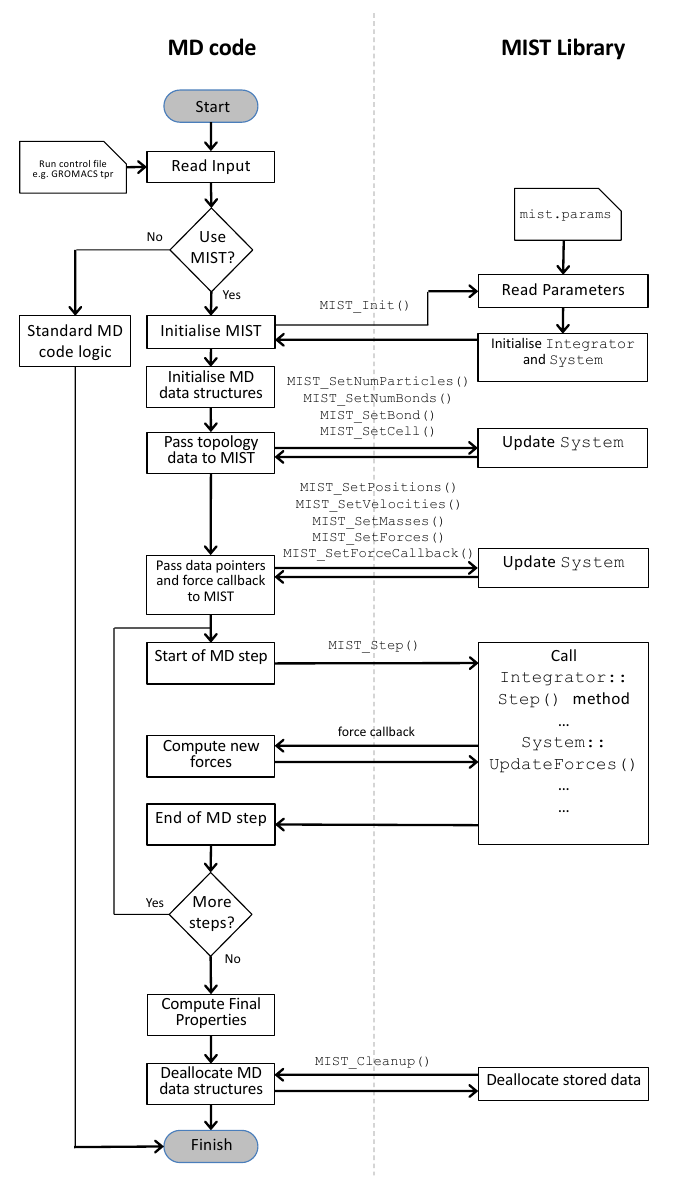}
\caption{Control flow in an MD code using the MIST library}
\label{fig:mist-flow}
}
\end{figure}

MD codes interact with the MIST library through a simple C or Fortran 90 API, which is designed to be general enough to interface to a wide range of possible MD codes.  The C interface is declared in a header \texttt{mist.h} and the Fortran interface in a module \texttt{mist\_f90}.  The Fortran interface contains exactly the same functionality as its C counterpart, with the only difference being that the functions are name \texttt{MIST\_F\_*} rather than \texttt{MIST\_*}.  As well as function declarations, a range of predefined constants (the aforementioned feature flags, and error codes) are also part of the interface.  MIST API calls may be added in to a host MD code either by hand, or for code versions that we support by automatically-applied source-code patches (the build process is explained in Section~\ref{sec:using-mist}).  MIST maintains an internal state and is responsible for its own memory management - a client calling the API always interacts with the same instance of the library, rather than for example having to pass an opaque handle back and forwards with each call.  Full documentation of the API is provided in Doxygen format, but the key concepts and ordering of API calls are shown in Figure~\ref{fig:mist-flow}, which outlines the typical control flow between a host MD code and MIST. 

It is important to note that by design, MIST extends the existing functionality of an MD code, and so if MIST is not selected as an option in the code's input configuration, it will behave exactly as normal.  Assuming that MIST has been selected, the first step is to initialise the library by calling \texttt{MIST\_Init()}.  This triggers MIST to read its own \texttt{mist.params} input file and initialise the \texttt{System} and \texttt{Integrator} objects accordingly.  If an error occurs at any stage (for example if the input file contained unrecognised keywords), the API returns an error code which the caller can check, and print an error message, or exit.  If the call was successful, MIST returns \texttt{MIST\_OK} (\texttt{0}).  Once the host code has completed initialisation to the point where the molecular topology is built and initial coordinates and velocities for the particles are assigned, this information must be passed to MIST.  A series of calls to (for example) \texttt{MIST\_SetNumParticles()}, \texttt{MIST\_SetNumBonds()} and \texttt{MIST\_SetPositions()} are used to inform MIST of the number of atoms and bonds, and to pass a pointer to the location of the particle position data.  In order to be completely flexible, data is passed across the API using void pointers, which are interpreted by MIST in the code-specific adaptor classes to yield data in the standard internal format provided by the \texttt{System} class for use by integrators.  The design decision to store pointers to the host data, enabling integrators to directly modify the simulation state, is chosen because it is more efficient that making an MIST-internal copy of the data, modifying that and explicitly copying it back before force updates, or the end of the MD step .

In addition to passing data pointers into MIST, the host MD code must also register a callback function pointer and associated parameters by calling \texttt{MIST\_SetForceCallback()}.  This callback function may be called by MIST during an MD step as a black-box to compute updated forces given the current atomic positions (and in general velocities).  Once again, we use a fully generic callback prototype, which accepts a single void pointer for any input data which may be required.  Since the actual force computation routine typically does not conform to this interface, it is convenient to define a lightweight parameter data type to store all the arguments which should be passed to the force computation routine and a wrapper function which unpacks the type and calls the appropriate function to compute updated forces.  For example, in GROMACS, we have:

\begin{verbatim}
typedef struct {
  FILE *log;
  t_commrec *cr;
  ...
  int flags;
  force_arrays_t *forces;
} force_params_t;

void do_force_wrapper(void *params){
    force_params_t *p = (force_params_t *)params;
    do_force(p->log,p->cr,p->inputrec,*(p->step),p->nrnb, \\
             *(p->wcycle),p->top,p->groups,*(p->box),p->x, \\
             p->hist,p->forces,*(p->vir_force),p->mdatoms, \\
             p->enerd,p->fcd,p->lambda,p->graph,p->fr \\
             p->vsite,*(p->mu_tot),p->t,p->field, \\
             p->ed,p->bBornRadii,p->flags);
}
...
// Declare a parameter type and store relevant local data in it
force_params_t p;
...
p.log = fplog;
p.cr = cr;
...
// Pass the function pointer and pointer to the parameter data to MIST
MIST_SetForceCallback(do_force_wrapper, &p)
\end{verbatim}

At this point MIST has all the data required to carry out a single MD step.  We note that, if for any reason the data pointers passed to MIST become out-of-date due to reallocation, or because the force parameters have changed, the \texttt{MIST\_Set\_*} functions must be called again as required.

In order to make using a MIST integrator as intuitive as possible for the user, we employ as much as possible of the unmodified code in the core MD time stepping loop.  In particular, if trajectory output and computation of thermodynamic variables such as temperature are done at the start of the step before a `native' integrator updates the system state, we do the same with MIST.  If they are done at the end, we do likewise.  However, in place of the native update code we insert a call to \texttt{MIST\_Step()}.  This hands over control to MIST to make whatever sequence of updates are implemented in the selected \texttt{Integrator}, including calls to the force callback routine as required during the step.  When \texttt{MIST\_Step()} returns, the system state has been advanced by a single time step \texttt{dt}, and any end-of-step actions which are required are taken such as incrementing step counters.  In our model, MIST is responsible only for the integration step itself.  This allows for a separation of concerns between configuring the integrator (via \texttt{mist.params}) and run control parameters e.g. number of steps, time step, output frequency and format, which are configured as usual for the host MD code.

Once the simulation has finished, the MIST library can be finalised by a call to \texttt{MIST\_Cleanup()}, which simply deallocates any memory which has been allocated to allow a clean shutdown of the MD code.

\subsection{Constraint Solver}
\label{sec:constraints}

In addition to the force-field i.e. the potential function of the atomic positions $U(\{\bm{r}\})$, from which the forces and therefore dynamics are derived, it is common in molecular simulations to apply \emph{constraints} to the system.  Bonds between heavy atoms and hydrogens have a high natural vibrational frequency and when using common integration algorithms such as the velocity Verlet method, which for a harmonic oscillator with frequency $\Omega$ has a stability threshold of $\delta t < 2/\Omega$, these vibrations severely limit the time step which may be used for stable MD (see \citep[Chapter~4.2]{Leimkuhler2015} for a more detailed discussion).  For applications such as conformational sampling, constraints are typically used to remove such vibrational degrees of freedom from the simulation, for example replacing flexible covalent bonds which are modelled as harmonic springs with rigid (fixed-length) `rods', thus allowing a larger time step and longer overall simulated time scales to be accessed for the same computational cost.  More complex constraints are also possible, including angular (fixing the internal angle between three atoms) and dihedral (fixing the torsional angle defined by four atoms), but these are not currently implemented in MIST.  For integrators to be practically useful, they must be able to generate a series of positions which satisfy the constraints, and so to avoid complicated coordinate transformation, additional steps are needed after the standard time-propagation of the positions and velocities to correct these back onto the \emph{constraint manifold} (the multidimensional surface made up of those points which satisfy the constraints).  These functions are provided by the MIST \texttt{ConstraintSolver} class and may be called by \texttt{Integrator}s.

As described in Section~\ref{sec:mist-system}, MIST has a representation of the molecular topology consisting of a set of bonds which link pairs of atoms $(a,b)$, with an equilibrium bond length $l$ (usually at the minimum of the bond-potential between the two atoms).  MIST supports applying constraints to three different groupings of bonds: none (\texttt{constraints off}), only bonds involving hydrogen atoms (\texttt{constraints h-bonds-only}), and all bonds (\texttt{constraints all-bonds}).  For the selected set of bonds, the \texttt{ConstraintSolver} sets up a list of $k$ \emph{holonomic} constraints (i.e. constraints depending only on the particle positions, and time), between atoms $ka,kb$ of the form:

\[
\sigma_k := ||\bm{r}_{ka} - \bm{r}_{kb}||^2 - l_k^2 = 0
\]

Following the standard approach~\cite{Rykaert1977} of considering the force $\bm{G}_i$ due to each constraint involving a particle $i$, which is defined by method of Lagrange multipliers as:

\[
\bm{G}_i = - \sum_k \lambda_k \nabla_i\sigma_k
\]

Then the constraints can be resolved (up to a defined tolerance) by solving for the Lagrange multipliers $\sigma_k$ and applying a correction to the unconstrained updated positions $\bm{\hat{r}}$:

\[
\bm{r}_i(t+\delta t) = \bm{\hat{r}}_i(t+\delta t) + \sum_k \lambda_k \frac{\partial \sigma_k}{\partial \bm{r}_i}
\]

By solving a second time for the set of Lagrange multipliers $\mu_k$ which satisfy the time derivative of the constraints:
%
%

\[
\frac{d \sigma_k(t)}{dt} = (\bm{v}_{ka} - \bm{v}_{kb})(\bm{r}_{ka} - \bm{r}_{kb}) = 0,
\]
where $\bm{v} = \dot{\bm{r}}$.

The unconstrained velocities $\bm{\hat{v}}$ may then be corrected by:

\[
\bm{v}_i(t+\delta t) = \bm{\hat{v}}_i(t+\delta t) + \sum_k \mu_k \frac{\partial \sigma_k}{\partial \bm{r}_i}
\]

Iterating through the constraints and adjusting the Lagrange multipliers, results in the RATTLE algorithm~\cite{Andersen1983}, and is selected with the keyword \texttt{constraints\_method rattle}.

MIST also implements the adaptive Symmetric Newton Iteration (SNIP) scheme~\cite{Barth1995}, where we construct a symmetric gradient matrix based on the configurations at the start of the timestep:
\[
\bm{\hat{R}} \equiv \sigma'(\{\bm{r}\})\bm{M}^{-1}\sigma'(\{\bm{r}\})^t
\]

Where $\sigma'(\{\bm{r}\})$ is the matrix of partial derivatives of the constraints with respect to the atomic coordinates and $\bm{M}$ is the diagonal matrix of particle masses.  Since the definition of a bond constraint involves only on a pair of atomic positions, the gradient matrix is sparse, with entries on the diagonal:

\[
\bm{\hat{R}}_{i,i} = ||\bm{r}_{ia} - \bm{r}_{ib}||^2 \left ( \frac{1}{m_{ia}} + \frac{1}{m_{ib}} \right ) = l_i^2 \left ( \frac{1}{m_{ia}} + \frac{1}{m_{ib}} \right )
\]

And off-diagonal for a pair of bonds $i,j$ where $ia = ja$ (and equivalent expressions for other combinations):

\[
\bm{\hat{R}}_{i,j} = \frac{(\bm{r}_{ia} - \bm{r}_{ib})(\bm{r}_{ia} - \bm{r}_{jb})}{m_{ia}}
\]

We can then solve for the set of Lagrange multipliers:

\[
\lambda_k(t) = \bm{\hat{R}}^{-1}\sigma_k(t)
\]

Update the positions:

\[
\bm{r}_i(t+\delta t) = \bm{\hat{r}}_i(t+\delta t) + \sum_k \lambda_k \frac{\partial \sigma_k}{\partial \bm{r}_i}
\]

And iterate these two steps until convergence. 

The velocity update is even simpler.  As in RATTLE we directly solve for the Lagrange multipliers which satisfy the time derivatives of the constraints:

\[
\mu_k(t) = \bm{\hat{R}}^{-1}\sigma'_k(t)
\]
And finally, set the velocities as for RATTLE. 

Importantly, for this method, the sparsity structure and the diagonal entries of the matrix $\bm{\hat{R}}$ are fixed for the duration of the simulation (since they depend only on the molecular topology), and the off-diagonal entries are fixed while the constraints are iterated.  In MIST, we make use of the Eigen library~\cite{eigenweb} to store the sparse matrix, perform a Cholesky factorisation, and solve for the Lagrange multipliers.  Eigen is particularly useful since we can perform a symbolic decomposition of the matrix once at the start of the simulation which makes the subsequent factorisation faster.  SNIP is the default constraint solution method in MIST.

\subsection{Units System}

One of the objectives of MIST was to make development of new integrators easy, and to enable a single implementation to be reused with multiple MD codes.  In addition to the abstractions discussed already, we must take account of the different units systems in use across different MD codes.  To avoid having to include code-specific scaling factors in the \texttt{Step()} function of individual integrators, we assume that the integration takes place using the same units system as the host code, and where there are parameters, we choose a unit and scale this into the internal units system using a series of convenience functions.  For example, in NAMD-Lite, energies are in kcal/mol, time is in femtoseconds, distances are in Angstroms.  To obtain a consistent units system, the internal mass unit is 0.0004184 amu i.e. a hydrogen atom has a `mass' of 2390.057... internal units.  Conversely, Amber uses a units system where masses are in amu, distance is in Angstroms, energies are in kcal/mol, and time is in units of 1/20.455 ps!

The \texttt{System} class provides functions which return standardised lengths (1 Angstrom), masses (1 amu), times (1 picosecond), and Boltzmann's constant in the internal units system of the host code - values which are provided by the code adaptor classes.  For example, in the Langevin dynamics integrator we require the constants $e^{-\gamma \delta t}$ and $\sqrt{k_B T (1 - e^{-2\gamma \delta t})}$. To convert the friction parameter $\gamma$ from the specified units of $ps^{-1}$ into internal time units we write:

\begin{verbatim}
double gdt = friction / system->Picosecond() * dt;
\end{verbatim}

Similarly, to get the Boltzmann factor $k_B T$ in internal energy units (T in Kelvin) and compute the two constants we can write:

\begin{verbatim}
double kbt = temp * system->Boltzmann();
...
double c1 = exp(-gdt);
double c3 = sqrt(kbt * (1 - c1 * c1));
\end{verbatim}

\section{Building and running MIST}
\label{sec:using-mist}

In order to use MIST with a particular MD code requires inserting MIST API calls into the source code and implementing a \texttt{System} adaptor class.  For codes which we provide adaptors for, source code patches are also given which can be applied during the build process.  To a user, the process is very simple:

\begin{itemize}

\item Configure MIST for use with a particular MD code, providing the location of the source code as an argument to the MIST \texttt{configure} script.  If configuration was successful, the script provides step-by-step instructions to complete the build for that specific MD code.  In general, the steps proceed as follows

\item Generate and apply the source code patch.  For the versions that we support (NAMD-Lite 2.0.3, GROMACS 5.0.2, Amber 14), the code patches will apply seamlessly.  It may be possible to apply the patches to other similar versions.

\item Build the MIST library.  Using a Makefile, MIST is built with support for the specified MD code compiled in.

\item Build the host code.  The host code is built (including the inserted MIST API calls) and linked with the MIST library.  Appropriate modifications to the build options are automatically made by the patching process, so no manual configuration such as library search paths or other linker flags is required from the user.

\end{itemize}

Once the MD code is built, it can be used entirely as normal.  To use a MIST integrator instead of the native ones provided by the host code requires adding a single parameter in the input file:

\begin{itemize}

\item NAMD-Lite: add \texttt{mist on} to the \texttt{.config} file.

\item GROMACS: set \texttt{integrator = mist} in the \texttt{.mdp} file.  Note that this change should be run through the GROMACS preprocessor \texttt{grompp} as usual, in order to have any effect.

\item Amber: set \texttt{imist = 1} in the \texttt{\&cntrl} namelist in the input file read by \texttt{pmemd}.

\end{itemize}

If MIST is enabled, execution will continue as described in Figure~\ref{fig:mist-flow} and timestepping will be controlled according to the \texttt{mist.params} file which is read by MIST, see Section~\ref{sec:integrators} for example parameters.  Any trajectory or diagnostic output options specified in the host code's input will still produce output in the usual format (MIST only modifies the dynamics).

\section{Performance Results}

We have measured the performance of MD simulations using MIST by comparing with the native integrators in each of the host codes we plug into.  CPU tests have been performed on ARCHER, a Cray XC30 with two Intel Xeon 12-core E5-2697v2 `Ivy Bridge' processors per node. We only use a single node due to the shared-memory parallelisation currently present in MIST.  GPU tests have been performed on a Linux system with two Intel Xeon 8-core E5-2650v2 `Ivy Bridge' processors and eight NVIDIA Tesla K40m GPUs.  GROMACS uses one GPU per MPI rank, so only a single GPU is used, even with multiple CPU threads.

All graphs show the average over 3 runs - error bars are not shown as the standard deviations were typically less than $<$ 1\%, with the largest being 2.5\%.  We used different test systems for each supported MD code, in order to avoid making direct comparison between the performance of the MD codes themselves, and also to illustrate the versatility of MIST to be able to cope with differing force fields, periodic boundary conditions and constraints schemes.

\subsection{NAMD-Lite}

\begin{figure}
\centering{
\includegraphics[width=4in]{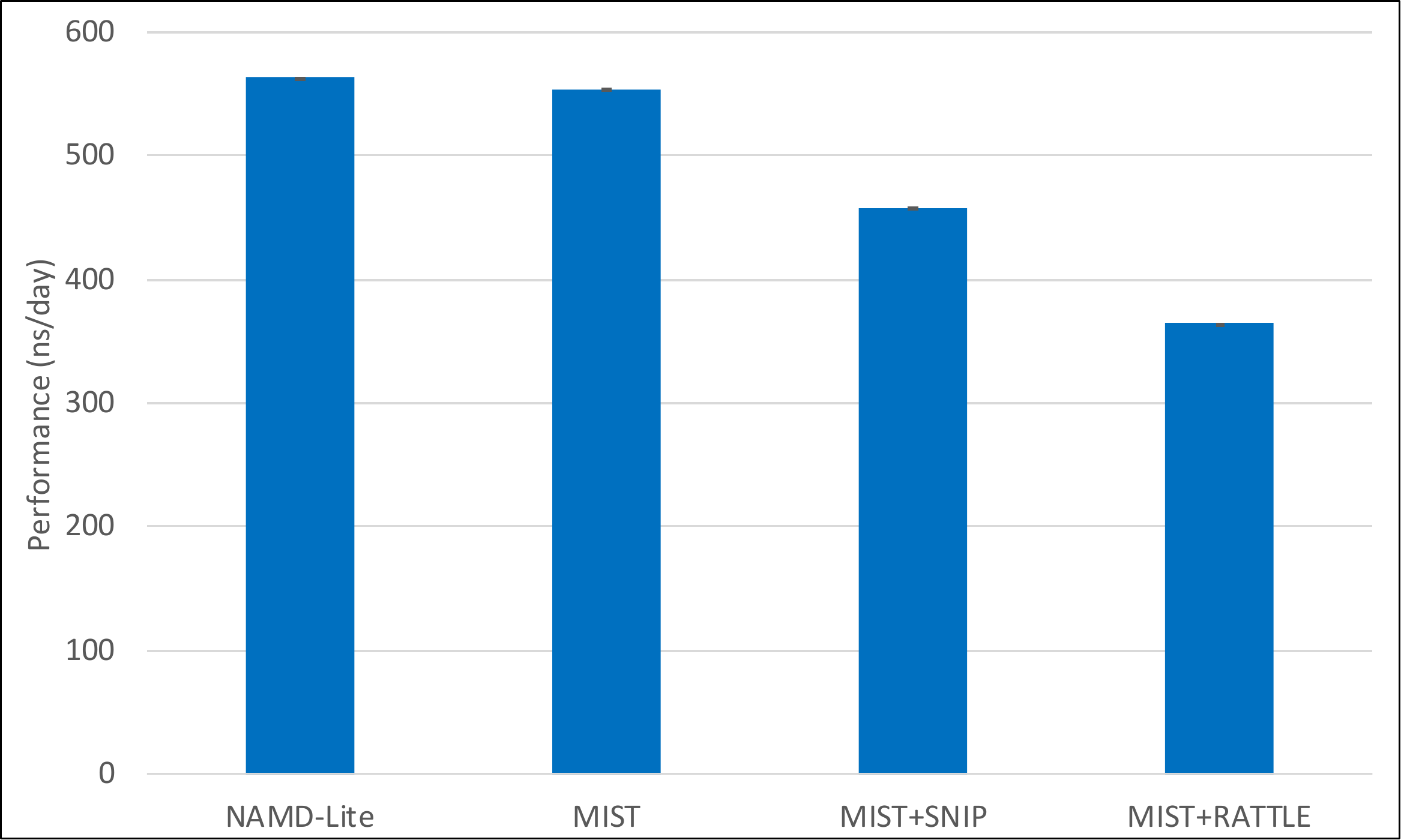}
\caption{Performance of NAMD-Lite with and without MIST, for a deca-alanine molecule \emph{in vacuo}.}
\label{fig:namd-lite-perf}
}
\end{figure}

As an initial test, we simulated an isolated deca-alanine molecule using the input settings supplied in the \texttt{demo} directory of the NAMD-Lite distribution (also in the \texttt{examples/namd-lite/alanin} directory of MIST).  A 12\AA\ cut-off is applied for electrostatic forces, the system is initialised with random velocities at 300K and we ran 1 ns of NVE dynamics using a 1 fs time step.  A small system is a worst-case test for MIST, since the relatively cheap force evaluation and small number of atoms (66) means that the function call overheads of calling out to MIST to perform the integration are likely to be exposed.

As can be seen in Figure~\ref{fig:namd-lite-perf}, we measured only a 2\% slowdown when using the velocity Verlet integrator in MIST compared with the native integrator in NAMD-Lite, effectively running with MIST disabled.  The figure shows the average over 3 runs for each setting - variability between runs was negligible (standard deviation $<$ 0.1\% in all cases).  The performance when constraints are enabled in MIST, using both the RATTLE~{\cite{Andersen1983}} and Symmetric Newton (SNIP)~{\cite{Barth1995}} methods, using the default constraint tolerance of $10^{-8}$ is shown.  Resolving the constraints takes a significant amount of time, with RATTLE reducing the overall performance by 34\%.  We find SNIP to be significantly faster, with only a 17\% performance drop.  We are not able to compare directly with NAMD-Lite's constraints implementation since it only supports the limited capability of SETTLE~\cite{Miyamoto1992} for rigid water models.

\subsection{GROMACS}

\begin{figure}
\centering{
\includegraphics[width=4in]{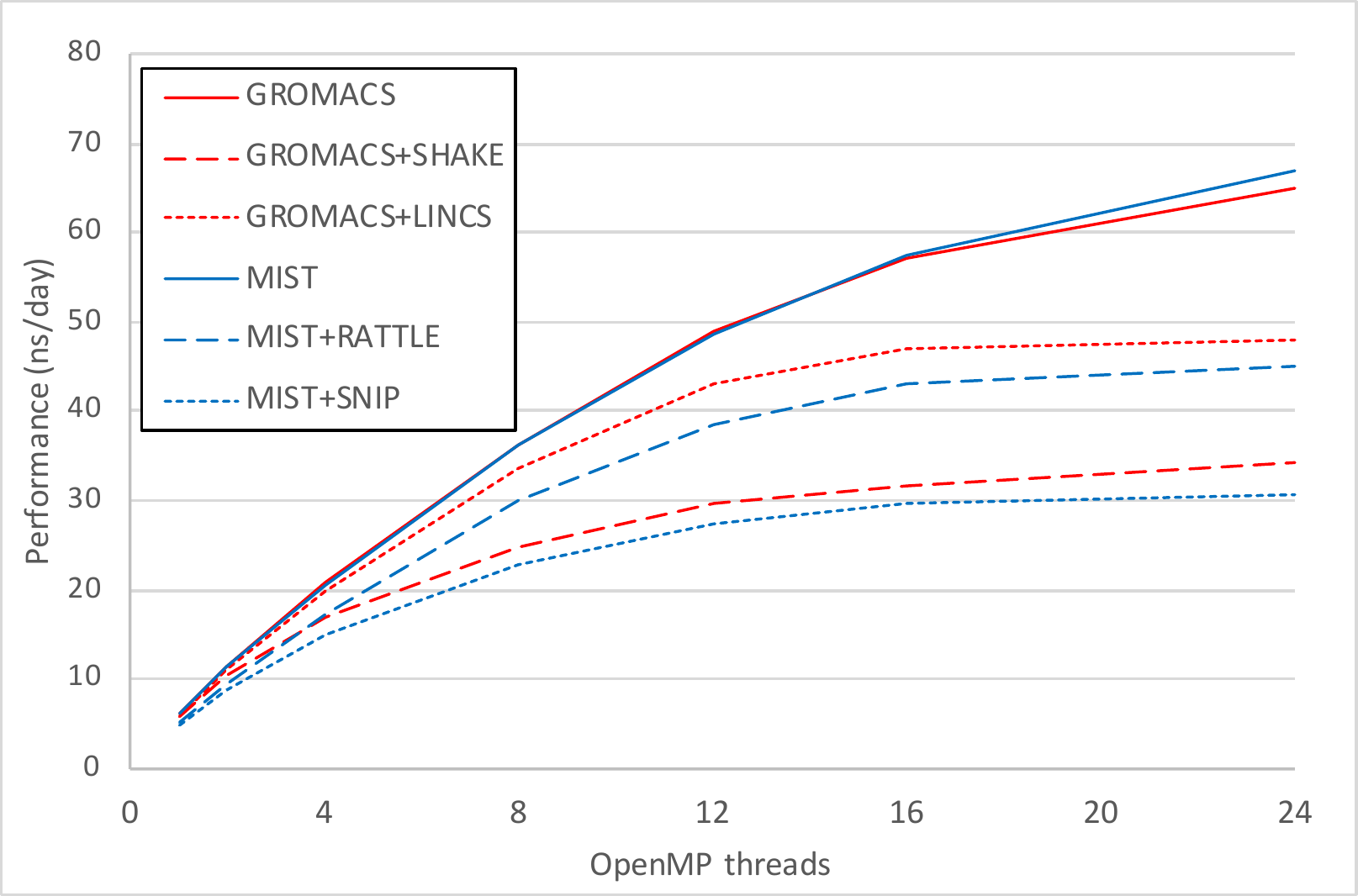}
\caption{Performance of GROMACS on ARCHER with and without MIST, for the 12,207 atom water system.}
\label{fig:gromacs-cpu-perf}
}
\end{figure}

\begin{figure}
\centering{
\includegraphics[width=4in]{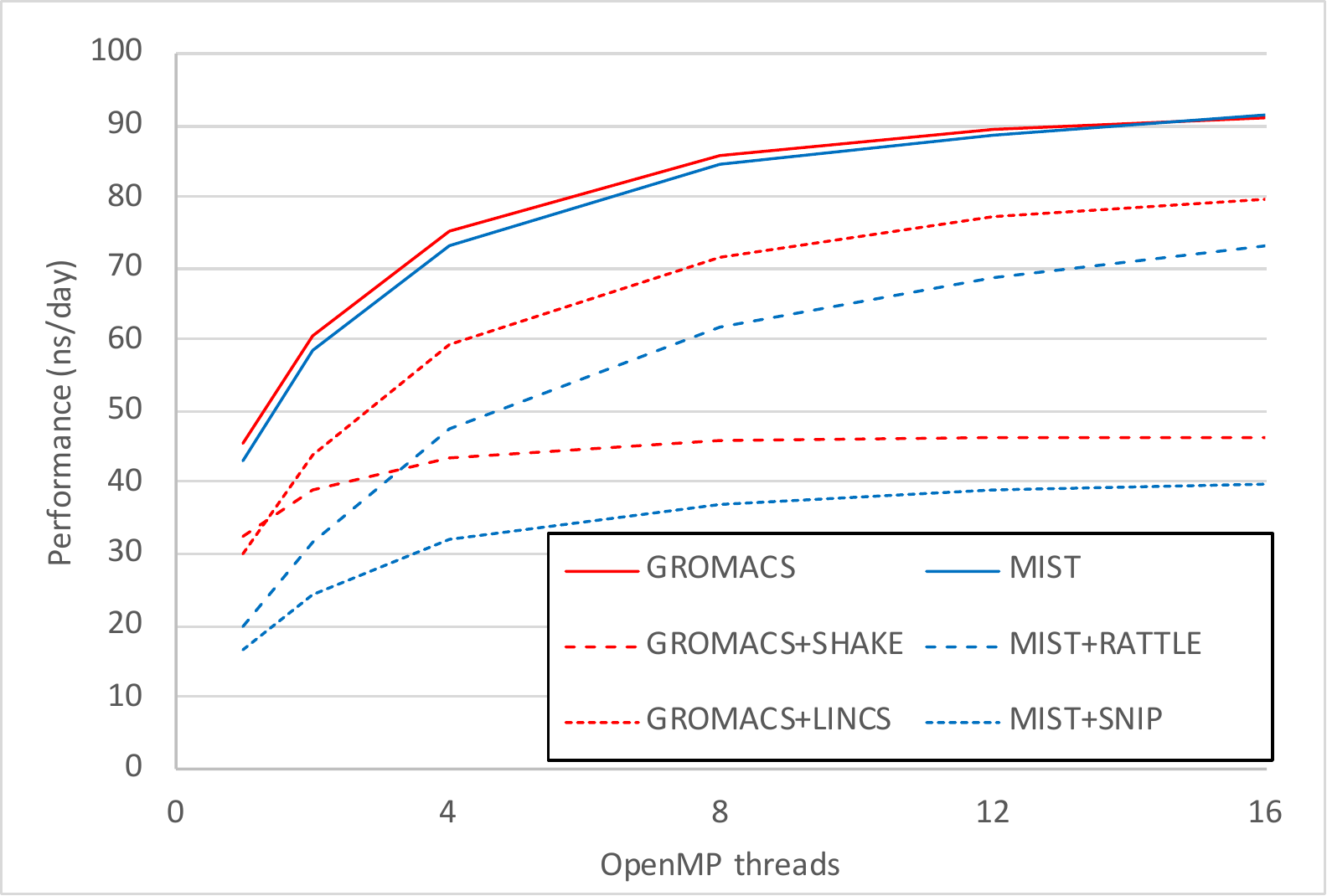}
\caption{Performance of GROMACS on an NVIDIA K40m GPU with and without MIST, for the 12,207 atom water system.}
\label{fig:gromacs-gpu-perf}
}
\end{figure}

We illustrate the performance of MIST with GROMACS using a more realistic-sized system - a 50\AA\ cubic periodic box containing 4069 water molecules.  The input geometry and settings are supplied in the \texttt{examples/gromacs/water} directory of the MIST distribution.  We use the TIP3P water model from the CHARMM27 force-field with a 12\AA\ cut-off for the electrostatic and van der Waals forces.  The system was initialized with random velocities at 300K and we ran 25 ps of NVE dynamics using a 1 fs time step.  For this system (using a single CPU core) 96\% of the run time is spent in force calculation and neighbor list search and less than 3\% in the integration itself (the \texttt{Update} time reported by GROMACS).  We tested both a fully flexible water model and one with bond constraints applied.  GROMACS supports the SHAKE and LINCS\cite{Hess1997} schemes for resolving constraints and we used the default settings for both.  In MIST, we used a relatively loose constraint tolerance of $10^{-4}$, matching the SHAKE tolerance used by GROMACS.

Figure~\ref{fig:gromacs-cpu-perf} shows the performance achieved for each case using up to 24 OpenMP threads.  For the unconstrained case, MIST is within 1\% of native GROMACS performance and on 24 threads outperforms GROMACS by 3\%.  Comparing the constraint implementations, the LINCS~{\cite{Hess1997}} algorithm in GROMACS performs best although it is not possible to directly compare it with the SHAKE, RATTLE or SNIP solvers as it does not use a relative constraint tolerance, but rather a fixed (4th) order expansion and a fixed number of iterations (1).  Interestingly, for the same tolerance, the RATTLE algorithm implemented in MIST is 12\% slower than GROMACS' SHAKE implementation when running on a single thread, but when using 24 threads is 36\% faster.  Unlike for the small deca-alanine system, SNIP is the slowest algorithm due to the expensive inversion of the gradient matrix.  It is important to recognize that SNIP is most advantageous when treating large macromolecules with high bond connectivity, whereas waters are actually better handled by RSHAKE~{\cite{Kol1997}} or SETTLE~{\cite{Miyamoto1992}}.

On the GPU (Figure~\ref{fig:gromacs-gpu-perf}), we see a slightly higher overhead between MIST and the native GROMACS Verlet integrator, of around 5.5\% using a single core, becoming negligible on 16 cores.  This reflects the fact that as the force evaluation using the GPU is much faster, with overall performance of 45 ns/day compared with 6 ns/day on the CPU only, the additional cost associate with calling out to MIST to do the integration is proportionally higher.  However, the use of OpenMP within the MIST integrator offsets this at higher thread counts.  Similarly to the CPU, the GROMACS LINCS implementation is fastest, but the difference in performance between the GROMACS SHAKE and MIST RATTLE implementations is much higher, with MIST being 39\% slower on a single thread, but 57\% faster using 16 threads.

\subsection{Amber}

\begin{figure}
\centering{
\includegraphics[width=4in]{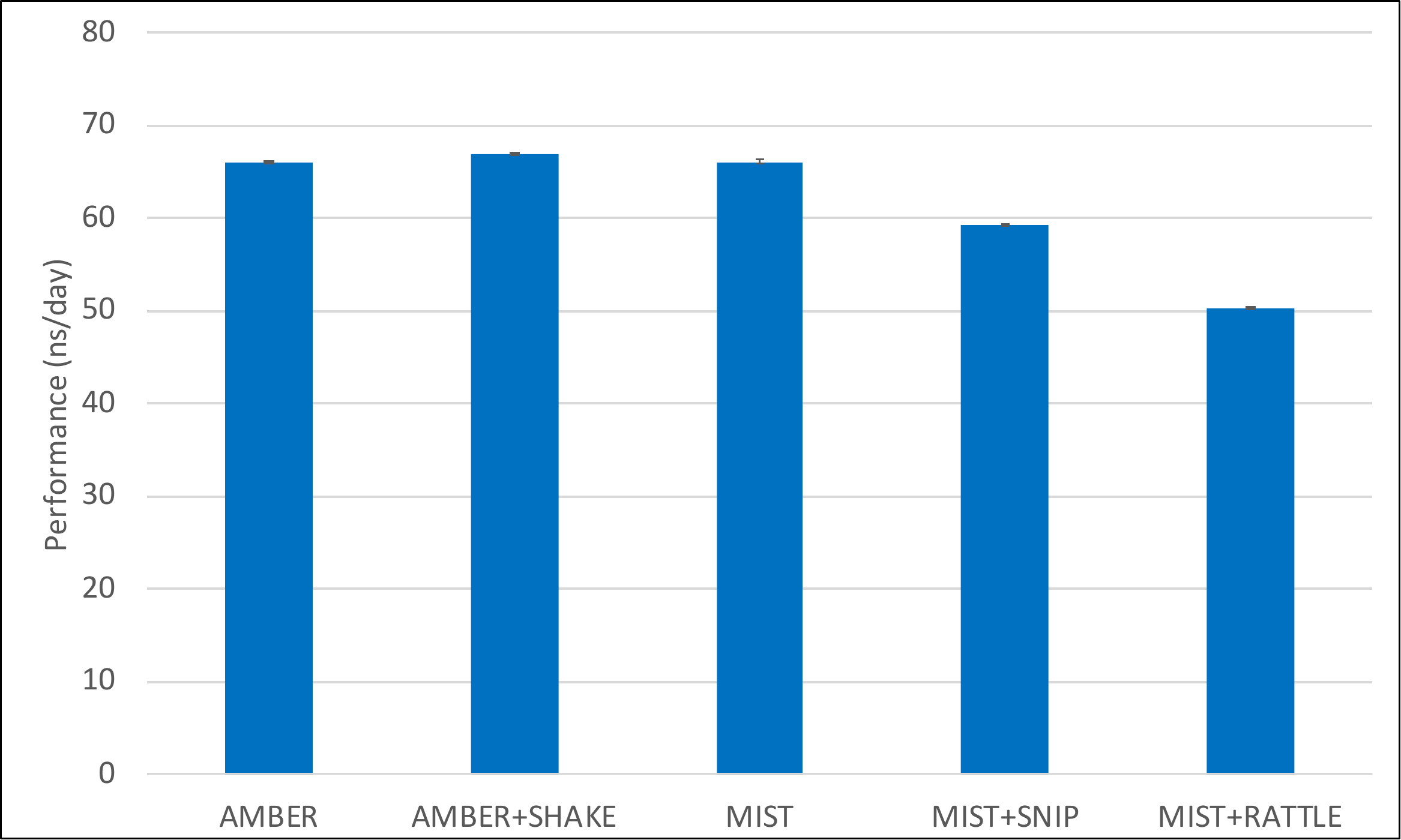}
\caption{Performance of Amber on ARCHER with and without MIST, for the solvated NTL9 system.}
\label{fig:amber-cpu-perf}
}
\end{figure}

\begin{figure}
\centering{
\includegraphics[width=4in]{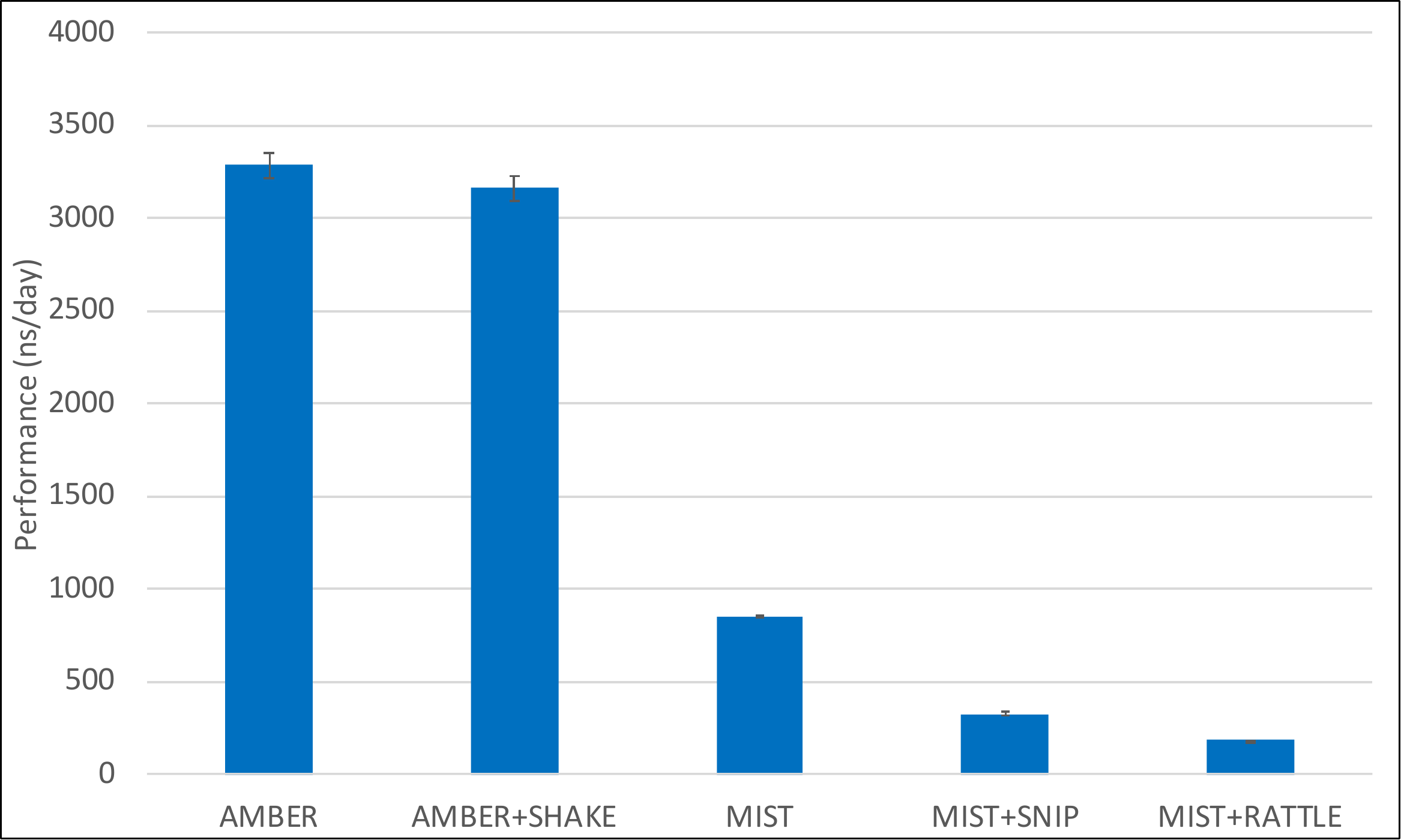}
\caption{Performance of Amber on an NVIDIA K40m GPU with and without MIST, for the solvated NTL9 system.}
\label{fig:amber-gpu-perf}
}
\end{figure}

To test the performance of MIST with Amber, we used the well-known~\cite{LindorffLarsen2011, Voelz2010} NTL9(1-39) protein, which consists of 636 atoms and is solvated in 4488 water molecules, for a total of 14100 atoms in an approximately 54\AA\ orthorhombic periodic unit cell.  The CHARMM22 force field with a TIP3P water model are used, with a cut-off of 9\AA\ for real-space part of the electrostatic forces and the long-ranged electrostatics computed on $54^3$ PME grid. The system was initialised with random velocities at 300K, and we ran 25ps of NVE dynamics using a 1 fs time step with the \texttt{pmemd} or \texttt{pmemd.cuda} program.  Input files are available in \texttt{examples/amber/ntl9} in the MIST distribution.  Amber supports bond constraints for hydrogen atoms only on the GPU, so we used the corresponding \texttt{h-bonds-only} setting in MIST, and a relative tolerance of $1E-5$, matching the default Amber SHAKE tolerance.

Figure~\ref{fig:amber-cpu-perf} shows the performance achieved running on a single CPU core on ARCHER (Amber 14 does not have thread parallelisation).  We see that using MIST has negligible impact on the performance.  For the constrained runs, we find that Amber is slightly faster (by 1\%) as it is possible to skip the computation of the forces caused by the constrained bonds (\texttt{ntf=2} in Amber), offsetting the additional cost of the SHAKE algorithm.  Similarly to NAMD-Lite and GROMACS, both the MIST constraint solver algorithms have an additional performance overhead.  For this system, SNIP is faster with an 11\% drop, compared to 24\% for RATTLE.

For Amber, the overhead of using MIST with \texttt{pmemd.cuda} is much higher (see Figure~\ref{fig:amber-gpu-perf}).  Whereas GROMACS achieves around $10\times$ speedup with a K40m GPU compared to a single CPU core, Amber achieves a speedup of over $50\times$ by doing the entire calculation (both the force evaluation and integration) on the GPU, thus avoiding relatively high-latency transfers between the GPU and CPU memory.  In order to use MIST the updated coordinates must be transferred to the GPU and the resulting forces transferred back again \emph{at each time step}, effectively throttling the GPU by memory transfers.  As a result, running with MIST achieves only 844 ns/day, compared with 3282 ns/day using the native integrator running on the GPU.  This is still over $10\times$ higher than the 66 ns/day achieved by Amber and/or MIST running on the CPU.  As expected, the constrained runs are slower still with a 62\% overhead for SNIP and 78\% overhead for RATTLE.  These overheads are higher than observed in the CPU runs as the force evaluation is much faster so the time spent in the constraint solver is proportionally larger. 

\subsection{Discussion}

The performance tests above show that, except in the case of Amber on GPU, performing integration within the MIST library has a negligible overhead compared to a native integrator.  Performing constrained integration comes at an additional overhead, the cost depending on the system size, topology and the constraint solver method and accuracy chosen, but is comparable to common methods such as SHAKE implemented in Amber and GROMACS.  As a result, MIST provides a suitable platform for developing new integrators with minimal loss of performance, but with much lower code complexity compared to developing an algorithm directly in one of the host codes.  In addition, we have shown that the same algorithms--implemented once in MIST--can be used in the context of 3 different codes, enabling greater applicability of any newly developed algorithm.

\section{Application: Simulated Tempering of Alanine-12}

Alanine-12 is a classic example of an $\alpha$-helical biomolecule.  It is particularly interesting to disentangle the effects of solvation and temperature on the unfolding process.  At room temperature, the unfolding process cannot be simulated directly with GROMACS because of slow kinetics.  An advanced sampling method such as Simulated Tempering could overcome this, but is not currently implemented in GROMACS.  To demonstrate the flexibility and capability of MIST, we implemented the Simulated Tempering algorithm of Nguyen \emph{et al} \cite{Nguyen2013,Zhang2015}, which previously have only been made available as a set of shell scripts \cite{ST}.  To set up a simulation using the scripts requires creating separate GROMACS input files for each temperature state, then running multiple short simulations, where the potential energy is parsed from the output file and a probabilistic change to another temperature state is made according to the algorithm.  As a result, the scripts generate a set of trajectory data files, which must be concatenated for analysis, and running many short individual simulations makes it inefficient to operate through an HPC batch system. An additional novel element of our implementation is that the temperature of each state is controlled using Langevin Dynamics implemented using a BAOAB splitting scheme \cite{Leimkuhler2013} to give more accurate configurational averages.  Our simulations are not designed to test this assertion, but a thorough analysis by Fass \emph{et al} {\cite{Fass2018}} showed dramatic reduction in configuration space discretisation error compared with other schemes.

To run Simulated Tempering through MIST requires a single long MD run, enabling MIST as described in Section \ref{sec:using-mist}.  The Simulated Tempering algorithm is selected and configured by a \texttt{mist.params} file as follows:

\begin{verbatim}
integrator simulated_tempering # Select Simulated Tempering
temperatures[0] 300            # Define a series of temperature states
temperatures[1] 310
temperatures[2] 320
...
temperatures[14] 440
temperatures[15] 450
period 2500                    # Attempt to switch states every 2500 steps
constraints all-bonds          # Apply bond constraints

langtemp 300                   # Start system at 300K
langfriction 1.0               # 1/ps friction constant
\end{verbatim}

\begin{figure}
\centering
\subfloat[NVT at 300K, starting from helical configuration]
{\includegraphics[width=2.5in]{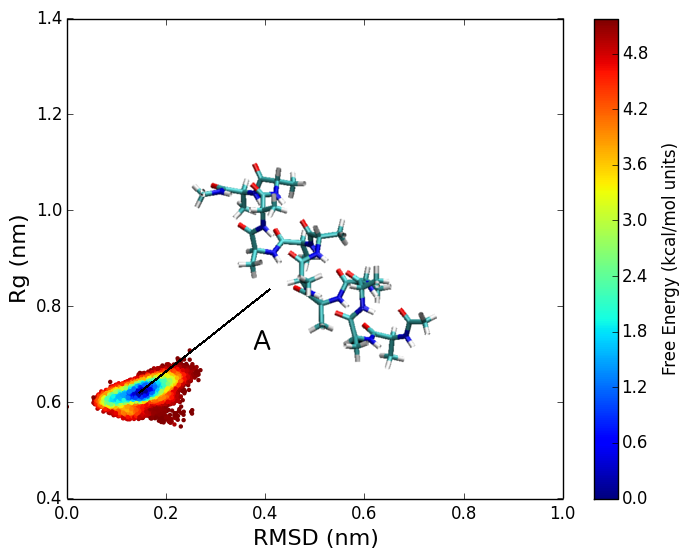}
\label{fig:ala-12-300-vac}
} \quad
\subfloat[Simulated Tempering 300-450K]
{\includegraphics[width=2.5in]{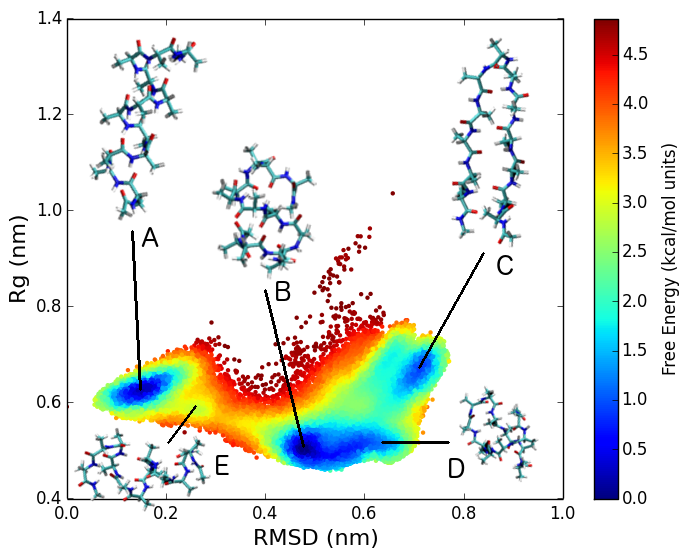}
\label{fig:ala-12-450-vac}
} \\
\subfloat[NVT at 300K, starting from configuration B]
{\includegraphics[width=2.5in]{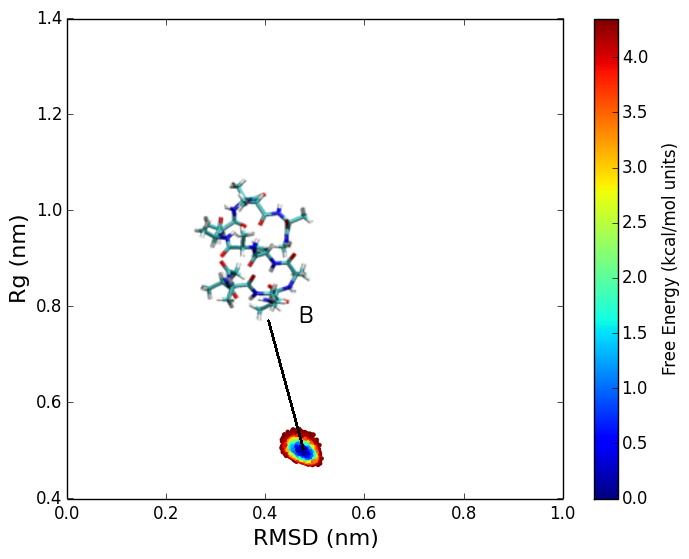}
\label{fig:ala-12-300-vac-B}
} \quad
\subfloat[NVT at 300K, starting from configuration C]
{\includegraphics[width=2.5in]{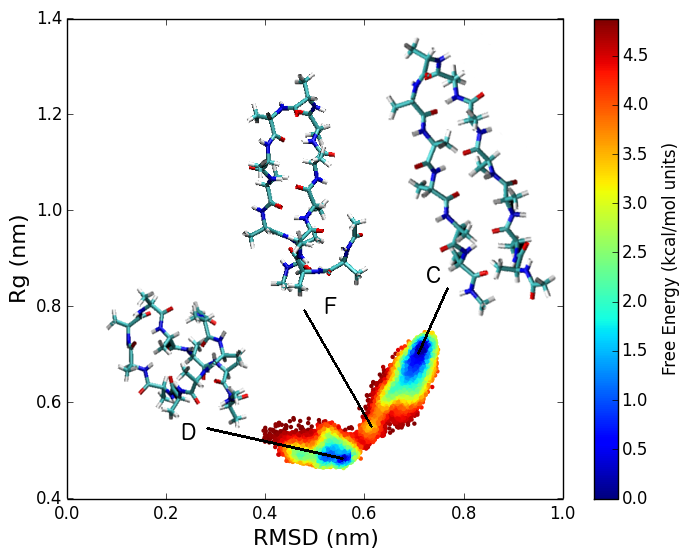}
\label{fig:ala-12-300-vac-E}
}
\caption{Free energy surfaces of Alanine-12 \emph{in vacuo}}
\label{fig:ala-12-vac}
\end{figure}

To test the implementation, we sampled the free energy landscape of the alanine-12 molecule, previously studied using the Diffusion-Map-directed-MD method \cite{Preto2014}.  Starting from the helical configuration \emph{in vacuo}, we ran a total of 1 $\mu s$ of MD using a 2 fs timestep.  We used the Amber96 forcefield with a 20\AA\ cut-off for electrostatics, constraining all bonds using the SNIP method.  Temperature was controlled using Langevin dynamics ($\gamma = 1.0 \mathit{ps}^{-1})$, either set to 300K to sample an NVT ensemble or varied using Simulated Tempering with temperature states ranging from 300K to 450K at 10K intervals.  The resulting free energy surfaces are plotted as a function of RMSD from the initial state and the radius of gyration $R_g$ in Figure \ref{fig:ala-12-vac}.

\begin{figure}
\subfloat[NVT at 300K, starting from helical configuration]
{\includegraphics[width=2.5in]{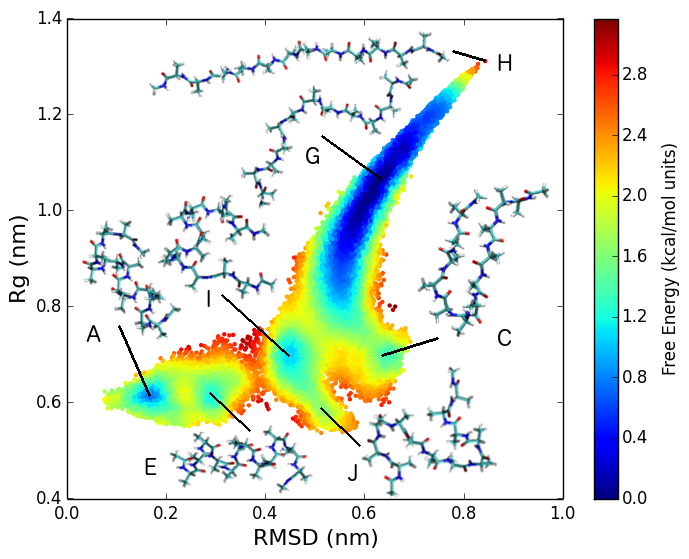}
\label{fig:ala-12-300}
} \quad
\subfloat[Simulated Tempering 300-450K]
{\includegraphics[width=2.5in]{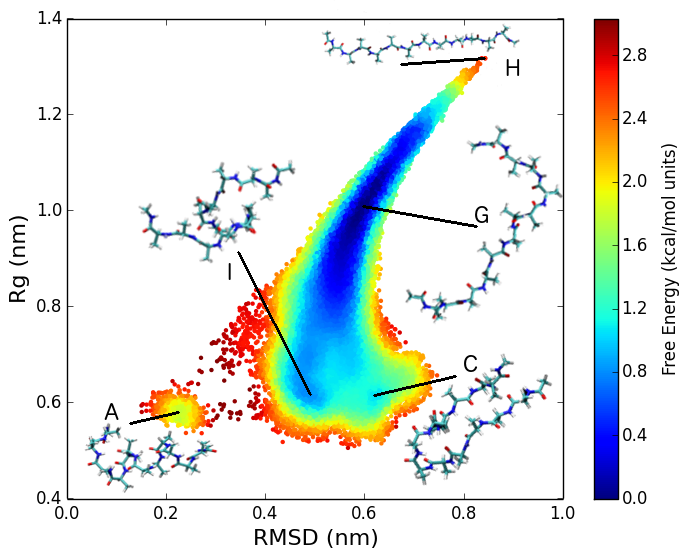}
\label{fig:ala-12-450}
}
\caption{Free energy surfaces of Alanine-12 solvated in TIP3P water.}
\label{fig:ala-12}
\end{figure}

As expected, Figure \ref{fig:ala-12-300-vac} shows that at 300K, the system is trapped in a local minimum around the helical state (labelled A).  Using Simulated Tempering, the elevated temperature is enough to allow the system to explore into a wider range of (partially uncoiled) configurations - comparable to those accessed by plain MD at 400K in Figure 7 of \cite{Preto2014}.  Figure \ref{fig:ala-12-450-vac} shows the complete set of configurations sampled, including those at temperatures greater than 300K.  Restarting simulations from the configurations labelled B (a compact structure consisting of three hairpin turns) and C (where the two termini are aligned and a complex twisted structure forms in the backbone) and running a subsequent 1 $\mu s$ of NVT dynamics at 300K shows that configuration B exists in a stable minima (Figure \ref{fig:ala-12-300-vac-B}), whereas the system is free to migrate between configurations C and D via a transition state F (Figure \ref{fig:ala-12-300-vac-E}), where the termini of the molecule have turned back on themselves.

In contrast, even at 300K (Figure \ref{fig:ala-12-300}) the solvated system is able to access a much wider range of states including the fully unfolded state (H) and a large basin (G) containing various extended structures.  This is qualitatively similar to the behaviour of deca-alanine observed in \cite{Hazel2014}, which has extended conformations of comparable free energy to the helical state.  Physically, the addition of water molecules provides an alternative hydrogen bonding route than can effectively `bridge' between -CO and -NH groups in the backbone, stabilising extended structures that are not observed in the \emph{in vacuo} ensemble.  Compared with the \emph{in vacuo} simulations, the molecule does not sample the compact ($Rg \simeq 0.5$) states B and D, but instead states like I and J where both ends of the molecule are unbound and a hairpin turn or complete helix is present in the middle.  Due to the relatively low energy barriers between states ($<2.8$kcal/mol) compared with barriers of up to 4 kcal/mol in the vacuum case, simulated tempering does not provide any access to any qualitatively new states (Figure \ref{fig:ala-12-450}).

\section{Conclusion and Future Development}
\label{sec:developments}

We have described the architecture and implementation of MIST, the Molecular Integration Simulation Toolkit, a C++ library which provides an abstraction layer over common MD codes to enable rapid development of new MD integration algorithms.  The initial release of the library contains implementations of six different integrators, and is interfaced via a C or Fortran API to four MD codes:  NAMD-Lite, GROMACS, Amber and Forcite.  MIST provides a portable platform for the development of novel integrators, which can be implemented once in MIST and used with any of the MD codes interfaced to MIST.  We have demonstrated the ease-of-use of MIST by implementing the simulated tempering scheme of Nguyen \emph{et al} \cite{Nguyen2013,Zhang2015} in combination with Langevin Dynamics using a `BAOAB' splitting \cite{Leimkuhler2013} and applying it to study the free energy landscape of Alanine-12 using GPU-accelerated GROMACS.  MIST is freely available under a BSD license from \url{https://bitbucket.org/extasy-project/mist}.

In serial, multi-threaded and GPU-accelerated configurations we have shown that MIST introduces negligible overhead compared to the native calculation, with the exception of Amber's GPU implementation.  In that case, the additional data transfer of the system state off the GPU introduces latency which slows the calculation down to performance comparable to GROMACS, where the native integration step is computed on the CPU and only forces are evaluated on the GPU.  It is possible to envisage a hybrid/multiple timestepping scheme where MIST is used to integrate the outer timestep for slow degrees of freedom such as a thermostat, and the inner timestep is integrated directly on the GPU using Amber's native integrator.  This has not yet been implemented, however.

At present, MIST is restricted to use in a shared-memory environment.  While it is possible to simulate quite large systems over reasonable timescales using multi-threading and/or GPU acceleration, to model very large macromolecules in complex environments requires the use of parallel computing using MPI.  Work is currently underway to remove the underlying assumption that the complete state of the system is available within a single address space, allowing MIST integrators to be used where the host MD code employs a domain decomposition.  The approach should be flexible enough to support particle-based, as well as space-partitioning strategies.  In addition, MIST itself needs to be parallelized, in particular the constraint solver, which must be able to resolve constraints potentially involving particles located on multiple processes.

While our initial release of MIST focuses on biomolecular applications, typically solvated molecules in an NVT ensemble, the tempering methods in particular are of interest for materials applications.  To represent a periodic solid, allowing for behavior such as expansion, contraction and phase transitions requires the lattice vectors of the simulation cell to become dynamical variables in the same sense as particle positions and velocities.  An interface to a materials simulation code such as LAMMPS \cite{Plimpton1995}, DL\_POLY \cite{Todorov2006}, or GULP \cite{Gale1997} combined with implementation in MIST of constant-pressure barostats such as Parrinello-Rahman \cite{Parrinello1980} or Martyna-Tuckerman-Klein \cite{Martyna1992} would open up usage of MIST to a wider user community.

\section*{Acknowledgements}

This work used the ARCHER UK National Supercomputing Service (\url{http://www.archer.ac.uk}) and systems at the STFC Hartree Centre (\url{http://www.hartree.stfc.ac.uk}).  Funding has been provided by the Engineering and Physical Sciences Research grants EP/K039512/1 and EP/P006175/1, the University of Edinburgh via a Staff Scholarship, the ERC grant HECATE and by the Hartree Centre.

We thank Charles Laughton for helpful suggestions on the interpretation of our Alanine-12 simulation results, and for the help of Phuong Nguyen on the implementation details of simulated tempering.


\section*{References}

\bibliographystyle{elsarticle-num}
\bibliography{mist}

\end{document}